\documentclass[12pt]{article}

\usepackage{amsfonts}
\usepackage{amsmath}
\usepackage{amsthm}
\usepackage{graphicx}
\usepackage{float}
\usepackage{multirow}
\usepackage[left=0.65in,right=0.65in,top=1in,bottom=1in]{geometry}
\numberwithin{equation}{section}
\begin{document}
\vspace{0.5in}

\begin{center}
{\LARGE{Anti-$\mathcal{PT}$-symmetric Qubit:\\
		 Decoherence and Entanglement Entropy}} 
\end{center}
\vspace{0.2in}
\begin{center}
{\large \bf Julia Cen and Avadh Saxena} \\ 
\vspace{0.2in}
{Theoretical Division and Center for Nonlinear Studies, 
Los Alamos National Laboratory, Los Alamos, New Mexico 87545, USA}\\
\vspace{0.05in}
{Emails: julia.cen@outlook.com, avadh@lanl.gov }
\end{center}
\vspace{0.2in}

{\bf {Abstract}}

We investigate a two-level spin system based anti-parity-time (anti-$\mathcal{PT}$)-symmetric qubit and study its decoherence as well as entanglement entropy properties. We compare our findings with that of the corresponding $\mathcal{PT}$-symmetric and Hermitian qubits. First we consider the time-dependent Dyson map to find the exact analytical dynamics for a  general non-Hermitian qubit system coupled with a bath, then we specialize it to the case of the anti-$\mathcal{PT}$-symmetric qubit. We find that the decoherence function for the anti-$\mathcal{PT}$-symmetric qubit decays slower than the $\mathcal{PT}$-symmetric and Hermitian qubits.  For the entanglement entropy we find that for the anti-$\mathcal{PT}$-symmetric qubit it grows more slowly compared to the $\mathcal{PT}$-symmetric and Hermitian qubits. Similarly, we find that the corresponding variance and area of Fisher information is much higher compared to the $\mathcal{PT}$-symmetric and Hermitian qubits.  These results demonstrate that anti-$\mathcal{PT}$-symmetric qubits may be better suited for quantum computing and quantum information processing applications than conventional Hermitian or even $\mathcal{PT}$-symmetric qubits.

\section{Introduction} 
One of the roadblocks for achieving viable quantum computing platforms is decoherence intrinsic to quantum systems \cite{Zurek2003}. This is particularly important for quantum information processing and storage \cite{Ladd2010}. Previously it was shown that parity-time or $\mathcal{PT}$-symmetric qubits are better than Hermitian qubits from this perspective when the $\mathcal{PT}$-symmetric quantum system is coupled to a Hermitian environment very weakly \cite{Gardas2016a}. Here we explore whether another non-Hermitian realization, i.e. an anti-$\mathcal{PT}$ qubit, can further improve decoherence properties. We find that the 
answer is in the affirmative. A very weak coupling to the bath or environment ascertains that the system and the bath do not exchange any heat \cite{Averin2016}, which leads to what is known as pure decoherence or dephasing \cite{Cummings2016}. The choice of a Hermitian bath is only for the sake of simplicity. 

\indent

Before exploring properties, we need to find the dynamics of our qubit systems. The computation of the qubit's reduced density matrix, through the expectation value of the time evolved operator, usually is performed from a Hermitian diagonal qubit, which can be easily obtained for Hermitian and $\mathcal{PT}$-symmetric qubits \cite{Gardas2016a,Luczka1990}. However, for the anti-$\mathcal{PT}$ qubit, we need to first provide a method to obtain the exact analytical expression for the evolved density matrix of a general non-Hermitian qubit system. This is accomplished through utilizing the time-dependent Dyson map to find a corresponding Hermitian system and as a result, obtain a reduced density matrix for a time-dependent Hermitian system \cite{Fring2019,Frith2020}. The decoherence is naturally revealed in the dynamics and we can subsequently also investigate further properties such as entanglement entropy (von Neumann as well as R\'{e}nyi) and Fisher information.

\indent

After the introduction of $\mathcal{PT}$-symmetry \cite{Bender1998} and subsequently two decades of intensive research \cite{Bender2019}, the notion of anti-$\mathcal{PT}$ symmetry was introduced by Ge and T\"ureci \cite{Ge2013} in optics by an appropriate spatial arrangement of the 
effective optical potential. The $\mathcal{PT}$ operator commutes with the Hamiltonian $\left[H,\mathcal{PT}\right]=0$  whereas the anti-$\mathcal{PT}$ operator anticommutes with the Hamiltonian $\{H, \mathcal{PT}\}=0$. The anti-$\mathcal{PT}$ symmetry has been realized in spatially coupled atom beams \cite{Peng2016}, electrical circuit resonators \cite{Choi2018}, optical waveguides with imaginary couplings 
\cite{Zhang2019} and optical four-wave mixing in cold atoms \cite{Jiang2019}. In addition, constant refraction optical 
systems \cite{Yang2017} and several experiments in atomic \cite{Wang2016, Chuang2018} and optical \cite{Konotop2018, 
qLi2019, Ke2019} systems have realized the anti-$\mathcal{PT}$ symmetry. There are many other applications involving waveguide 
arrays  \cite{Ke2019}, diffusive systems \cite{Li2019}, phase transitions \cite{Lee2014}, spin chains \cite{Couvreur2017}, 
information flow \cite{Chakraborty2019} and non-Markovian aspects \cite{Haseli2014}. 

\indent

Specifically, we investigate an anti-$\mathcal{PT}$-symmetric quantum system very weakly coupled to a bath (or environment). This leads to a critical slowing down of decoherence that is better than its $\mathcal{PT}$-symmetric counterpart.  Possibly, an anti-$\mathcal{PT}$ qubit can be experimentally realized in recently demonstrated optical and microcavity settings \cite{Zhang2020}. A quantum circuit \cite{Zheng2019} and information flow \cite{Wen2020} using a two-level system have also been recently discussed. 

\indent

The outline of the paper is as follows. We first introduce the two-level Hermitian, $\mathcal{PT}$-symmetric and anti-$\mathcal{PT}$-symmetric qubits.
For the former two qubits, we explicitly write down their reduced density matrices. In Section 2 we show a general formalism for treating non-Hermitian qubits by way of the time-dependent Dyson map for the density matrix and apply it to the anti-$\mathcal{PT}$-symmetric qubit. Section 3 is devoted to studying decoherence of the anti-$\mathcal{PT}$ qubit. In Section 4 we calculate the von Neumann and R\'{e}nyi entanglement entropy for the anti-$\mathcal{PT}$ qubit. In Section 5 we invoke the use of the Kullback-Leibler divergence in order to compute the quantum Fisher information \cite{Kullback1978, Bao2020}.  Section 6 deals with spin vector representation for the three types of qubits. Finally, Section 7 contains a summary of our main conclusions. 

\subsection{The Model}

In this paper, we will be considering the spin qubit system coupled to a bath of bosonic systems described by the following Hamiltonian \cite{Breuer2002}
\begin{equation}
	H=H_{S}\otimes\mathbb{I}_{B}+\mathbb{I}_{S}\otimes H_{B}+H_{S}\otimes V_{B},
\end{equation}
where $H_{S}$ denotes the system, $H_{B}$ the bath
\begin{equation}
	H_{B}=\sum_{k} \omega_{k}a^{\dagger}_{k}a_{k},
\end{equation}
and $V_{B}$ the interaction term between the system and bath
\begin{equation}
	V_{B}=\sum_{k} \left(g_{k}a^{\dagger}_{k}+g^{\ast}_{k}a_{k}\right).
\end{equation}
Here $a_k^\dagger$ and $a_k$ denote bosonic creation and annihilation operators, respectively, $\omega_{k}$ are the eigenmodes of the bath and $g_k$ are the coupling constants. One of the main purposes of this paper is to compare the various features and properties between a Hermitian, $\mathcal{PT}$-symmetric and non-Hermitian qubit system, with particular focus on the anti-$\mathcal{PT}$-symmetric qubit. 

\subsection{Hermitian Qubit} 

To begin, let us review the dynamics for the Hermitian qubit as
\begin{equation}
	H^{h}_{S}=\begin{pmatrix}
	\alpha+\theta & \xi+i\delta \\ 
	 \xi-i\delta & -\alpha+\theta
	\end{pmatrix},\label{HQ}
\end{equation}
with $\alpha,\xi,\delta,\theta\in\mathbb{R}$ to satisfy the Hermiticity condition $H^{h}_{S}=\left(H^{h}_{S}\right)^{\dagger}$. For convenience, we take a similarity transformation \cite{Strang2006}
\begin{equation}
	T=\begin{pmatrix}
		\omega_{0}-\alpha & -\xi-i\delta \\ 
		\omega_{0}+\alpha & \xi +i\delta
		\end{pmatrix}, \quad \text{where} \quad \omega_{0}=\sqrt{\alpha^{2}+\delta^{2}+\xi^{2}},\label{T}
\end{equation}
to obtain a diagonalized Hamiltonian $H^{Dh}_{S}=TH^{h}_{S}T^{-1}$. Subsequently, the evolved reduced density matrix for the system can be expressed in terms of the reduced density matrix for the diagonalized system with normalization as
\begin{equation}
	\rho_{S}\left(t\right)=\frac{T^{-1}\rho_{S}^{Dh}\left(t\right)\left(T^{-1}\right)^{\dagger}}{Tr_{S}\left[T^{-1}\rho_{S}^{Dh}\left(0\right)\left(T^{-1}\right)^{\dagger}\right]}
\end{equation}
and the diagonalized reduced density matrix being
\begin{equation}
	\rho_{S}^{Dh}\left(t\right)=\begin{pmatrix}
							\rho_{11}^{Dh} & 0 \\ 
							0 & \rho_{22}^{Dh}
							\end{pmatrix}+
							\begin{pmatrix}
							0 & \rho_{12}^{Dh}e^{2i\omega_{0}t}e^{-i\omega_{0}\Omega\left(t\right)} \\ 
							\rho_{21}^{Dh}e^{-2i\omega_{0}t}e^{i\omega_{0}\Omega\left(t\right)} & 0
							\end{pmatrix}e^{-\omega_{0}^{2}\gamma\left(t\right)},\label{HermitianrhoDS}
\end{equation}
where
\begin{align}
	\Omega\left(t\right)&=4\theta\int_{0}^{\infty}dw J\left(w\right)\frac{wt-\sin\left(wt\right)}{w^{2}},\\
	\gamma\left(t\right)&=4\int_{0}^{\infty}dw J\left(w\right)\frac{1-\cos\left(wt\right)}{w^{2}}\coth\left(\frac{\beta w}{2}\right),\\
	J\left(w\right)&=\sum_{k}\lvert g_{k}\rvert^{2}\delta\left(w-w_{k}\right) \,, 
\end{align}
are respectively the function influencing phase evolution, decoherence function and spectral density of the bath. 

\subsection{$\mathcal{PT}$-symmetric Qubit}

For the $\mathcal{PT}$-symmetric case, which has been studied in \cite{Gardas2016a}, let us take the Hamiltonian as
\begin{equation}
	H^{\mathcal{PT}}_{S}=\begin{pmatrix}
		\alpha+i \theta & \xi+i\delta \\ 
		\xi-i\delta & \alpha-i \theta
		\end{pmatrix},\label{PTQ}
\end{equation}
where $\alpha,\xi,\delta,\theta\in\mathbb{R}$, $\delta^{2}+\xi^{2}\geq \theta^{2}$ and the parity and time operators to be
\begin{equation}
	\mathcal{P}=\sigma_{x}=\begin{pmatrix}
				0 & 1 \\ 
				1 & 0
				\end{pmatrix}, \qquad \mathcal{T}: i\rightarrow -i,
				\label{PT}
\end{equation}
which has the property $\left[\mathcal{PT},H^{\mathcal{PT}}_{S}\right]=0$. Taking the same similarity transformation (\ref{T}) as for the Hermitian system, but where
\begin{equation}
	 \omega_{0}=\sqrt{\gamma^{2}+\delta^{2}-\theta^{2}},
\end{equation}
will lead to a diagonalized Hermitian Hamiltonian $h^{\mathcal{PT}}_{S}=TH^{\mathcal{PT}}_{S}T^{-1}$ and same evolved reduced density matrix (\ref{HermitianrhoDS}), with the difference that $\Omega\left(t\right)$ is now $-\Omega\left(t\right)$.

\subsection{Anti-$\mathcal{PT}$-symmetric Qubit}

We now introduce an anti-$\mathcal{PT}$-symmetric quantum system of the general form
\begin{equation}
	H_{S}= \begin{pmatrix}
		\alpha+i\theta & \xi+i\delta \\ 
		-\xi+i\delta & -\alpha+i\theta
		\end{pmatrix},\label{APTQ}
\end{equation}
where $\alpha,\xi,\delta,\theta\in\mathbb{R}$ and can check that the anticommutation relation $\left\{\mathcal{PT},H_{S}\right\}=0$ is satisfied taking the same parity and time operators (\ref{PT}). The total system $H$, is diagonalizable by again taking the similarity transformation (\ref{T}) with
\begin{equation}
\omega_{0}=\sqrt{\alpha^{2}-\xi^{2}-\delta^{2}}.
\end{equation}
The resulting diagonalized Hamiltonian will be given by
\begin{alignat}{1}
H^{D}&=THT^{-1}, \label{HD}\\
&=\left(-\omega_{0}\sigma_{z}+i\theta\mathbb{I}_{S}\right)\otimes\mathbb{I}_{B}+\mathbb{I}_{S}\otimes H_{B}+\left(-\omega_{0}\sigma_{z}+i\theta\mathbb{I}_{S}\right)\otimes V_{B}. \notag
\end{alignat}
This is equivalent to the eigenbasis representation   
\begin{equation}
	THT^{-1}=\sum_{n} E_{n}\lvert n\rangle\langle n\rvert
\end{equation}
with $E_{n}\in \mathbb{C}$, which can be rewritten in a complete biorthonormal basis for $H$ \cite{Wong1967,Wong1969,Mostafazadeh2002,Brody2013}
\begin{align}
	H&=T^{-1}\sum_{n}E_{N}\lvert n\rangle\langle n\rvert T, \\
	&=\sum_{n}E_{N}\lvert\psi_{n}^{R}\rangle\langle\psi_{n}^{L}\rvert, \notag
\end{align}
taking the set of eigenvectors $\lvert\psi_{n}^{R}\rangle=T^{-1}\lvert n\rangle$ and $\langle\psi_{n}^{L}\rvert=\langle n \rvert T$, satisfying the defining equations $\langle\psi_{n}^{L}\vert\psi_{m}^{R}\rangle=\delta_{nm}$ and $\sum_{n}\lvert\psi_{n}^{R}\rangle\langle\psi_{n}^{L}\rvert=\mathbb{I}$. 

\indent

The pair of eigenvalues of the anti-$\mathcal{PT}$-symmetric system (\ref{APTQ}) is given by $E_{\pm}=i\theta\pm\omega_{0}$ with the difference of the two eigenvalues being $\widetilde{E}=2\omega_{0}$. This represents the energy gap of our two-level system and its reality depends on the parameters $\alpha$, $\delta$ and $\xi$. When $\delta^{2}+\xi^{2}=\alpha^{2}$, the energy gap will be zero to give degenerate eigenvalues, but for this paper, we will be interested in the case when the energy gap is real, i.e. to study the parametric domain $\alpha^{2}\geq\delta^{2}+\xi^{2}$.

\indent

For a general non-Hermitian Hamiltonian, $H$, it has been suggested in \cite{Brody2012, Sergi2013, Wang2020} that the Hamiltonian can also be viewed as a decomposition of real and imaginary parts $H=H_{R}+iH_{I}$ to give a complex extension of the Liouville-von Neumann equation
\begin{equation}
	\rho_{t}=-i\left[H_{R},\rho\right]+\left\{H_{I},\rho\right\}-2\rho Tr\left(\rho H_{I}\right),
\end{equation}
which can be solved by the form
\begin{equation}
	\rho\left(t\right)=\frac{U\left(t\right)\rho\left(0\right)U^{\dagger}\left( t\right)}{Tr \left[U\left(t\right)\rho\left(0\right)U^{\dagger}\left(t\right)\right]},
\end{equation}
given the evolution operator $U\left(t\right)=e^{-iHt}$, then tracing out the bath degrees of freedom results in the reduced density matrix for the system. However, for non-Hermitian spin-boson models, calculations become quite involved. In what follows, we will present a scheme that makes computing the reduced density matrix of these systems more feasible.  

\section{Time-dependent Dyson Map for Density Matrix of a Non-Hermitian System}

The key is to show that one can reformulate the non-Hermitian Hamiltonian problem in terms of a Hermitian one utilizing a time-dependent Dyson map \cite{Faria2006a,Mostafazadeh2007,Znojil2008,Gong2013,Fring2016,Fring2017a,Fring2017b,Mostafazadeh2018,Cen2019}. Recently, this has been investigated for a $\mathcal{PT}$-symmetric bosonic system coupled to a
bath of N bosonic systems \cite{Fring2019} and $\mathcal{PT}$-symmetric Jaynes-Cummings Hamiltonian \cite{Frith2020}. Here, we shall present this method for a general non-Hermitian qubit system and in particular we will focus on the anti-$\mathcal{PT}$-symmetric case.

\indent

First, let us recall that for a Hermitian system with density matrix $\rho^{Dh}$, the Liouville-von Neumann equation is given by 
\begin{equation}
	i\rho_{t}^{Dh}=\left[h^{D},\rho^{Dh}\right].\label{Liouvilleh}
\end{equation}
In the non-Hermitian case, taking the diagonalized example $H^{D}$, we can write the Schr\"{o}dinger equation and its conjugate transpose as 
\begin{align}
	i\frac{\partial}{\partial t}\lvert \psi_{i}\rangle&=H^{D}\lvert \psi_{i}\rangle,\\
	-i\frac{\partial}{\partial t}\langle \psi_{i}\rvert&=\langle \psi_{i}\rvert \left(H^{D}\right)^{\dagger},
\end{align}
where, with $\rho^{D}=\sum_{i}P_{i}\lvert \psi_{i}\rangle\langle \psi_{i}\rvert$ being the density matrix, the corresponding non-Hermitian Liouville-von Neumann equation can be derived as
\begin{align}
	i\rho^{D}_{t}&=i\sum_{i}P_{i}\left(\frac{\partial}{\partial t}\lvert \psi_{i}\rangle\langle \psi_{i}\rvert+\lvert \psi_{i}\rangle\frac{\partial}{\partial t}\langle \psi_{i}\rvert\right),\\
	&=H^{D}\rho^{D}-\rho^{D}\left(H^{D}\right)^{\dagger}. \label{LiouvilleHD}
\end{align}
Let $H^{D}$ and $h^{D}$ be related by the Dyson relation 
\begin{equation}
	H^{D}=\eta^{-1}h^{D}\eta-i\eta^{-1}\eta_{t},\label{dyson}
\end{equation}
where $\eta$ relates the states $\lvert \phi_{i}\rangle$, $\lvert \psi_{i}\rangle$ of the Hamiltonians $h^{D}$, $H^{D}$ respectively as $\lvert \phi_{i}\rangle=\eta\lvert \psi_{i}\rangle$. Substituting the Dyson relation into (\ref{LiouvilleHD}) and comparing with (\ref{Liouvilleh}) gives the relation of density matrices between the Hermitian and non-Hermitian Hamiltonians as
\begin{equation}
	\rho^{Dh}=\eta\rho^{D}\eta^{\dagger}.\label{dysondensity}
\end{equation}
Supposing $\rho^{Dh}=\sum_{i}P_{i}\lvert\phi_{i}\rangle\langle\phi_{i}\rvert$, we can check using the relation above that $\rho^{D}=\sum_{i}P_{i}\lvert\psi_{i}\rangle\langle\psi_{i}\rvert$, which shows the mapping of density matrices is able to preserve the set of probabilities $P_{i}$.

\indent

On the other hand, taking the Hermitian Liouville-von Neumann equation (\ref{Liouvilleh}) and substituting again the Dyson relation (\ref{dyson}), it can be shown
\begin{equation}
	i\widetilde{\rho}_{t}^{D}=\left[H^{D},\widetilde{\rho}^{D}\right]
\end{equation}
under the relation $\widetilde{\rho}^{D}=\eta^{-1}\rho^{Dh}\eta=\rho^{D}M$, with $M=\eta^{\dagger}\eta$ being the metric such that for the quasi-Hermitian Hamiltonian
\begin{equation}
	H^{Q}=H^{D}+i\eta^{-1}\eta_{t},
\end{equation}
it satisfies $\langle H^{Q}\psi_{i}\vert M\psi_{i}\rangle=\langle\psi_{i}\lvert MH^{Q}\psi_{i}\rangle=\langle\phi_{i}\vert h^{D}\phi_{i}\rangle$ and $\left(H^{Q}\right)^{\dagger}M=MH^{Q}$. Looking at the anti-$\mathcal{PT}$-symmetric qubit of (\ref{HD}) which we now denote by $H^{D}_{S}$, we will see that the corresponding quasi-Hermitian qubit $H^{Q}_{S}$, is a two-level system with energies $E_{\pm}^{Q}=\pm \omega_{0}$, with same energy gap as $H^{D}_{S}$. It follows that $H^{Q}_{S}$ is interpreted as the {\it physical operator} that plays the role of energy for $H^{D}_{S}$ \cite{Fring2016,Fring2017a}, and $\widetilde{\rho}^{D}$ is a Hermitian density matrix operator in the Hilbert space under the metric $M\left(t\right)$. We now find the corresponding Hermitian system from a time-dependent Dyson map. 

\indent

Let us recall the diagonalized non-Hermitian Hamiltonian (\ref{HD}), if we take the Ansatz $\eta=e^{\epsilon\left(t\right)}e^{\varphi\left(t\right)V_{B}}$ in the Dyson relation, $h^{D}$ becomes
\begin{equation}
	h^{D}=-\omega_{0}\sigma_{z}+i\theta+H_{B}-\varphi\left(t\right)\widetilde{V}_{B}-\varphi^{2}\left(t\right)\Omega_{k}-\omega_{0}\sigma_{z}V_{B}+i\theta V_{B}+i\dot{\epsilon}\left(t\right)+i\dot{\varphi}\left(t\right)V_{B},
\end{equation}
where $\widetilde{V}_{B}=\sum_{k}\omega_{k}\left(g_{k}a^{\dagger}_{k}-g_{k}^{\ast}a_{k}\right)$ and $\Omega_{k}=\sum_{k}\omega_{k}\lvert g_{k}\rvert^{2}$. For $h^{D}$ to be Hermitian $\left(h^{D}\right)^{\dagger}=h^{D}$, the constraining equations are $\dot{\epsilon}\left(t\right)=-\theta$ and $\dot{\varphi}\left(t\right)=-\theta$, so we can take $\eta=e^{-\theta t}e^{-\theta tV_{B}}$, then the corresponding Hermitian Hamiltonian is
\begin{equation}
	h^{D}=-\omega_{0}\sigma_{z}+H_{B}-\omega_{0}\sigma_{z}V_{B}+\theta t\widetilde{V}_{B}-\theta^{2}t^{2}\Omega_{k}. \label{hD}
\end{equation}

\indent

Note we can also represent $h^{D}$ in terms of the quasi-Hermitian Hamiltonian $H^{Q}$ \cite{Mostafazadeh2005, Faria2006b, Gardas2016b}, as
\begin{equation}
	h^{D}=\sum_{n=0}^{\infty}\frac{\left(-1\right)^{n}}{n!}C_{G}^{\left(n\right)}\left(H^{Q}\right),
\end{equation}
with $G=\theta\left(1+V_{B}t\right)$ and denoting
\begin{equation}
	C_{G}^{\left(n\right)}\left(\mathcal{O}\right)=[\overbrace{G,[G,\cdots[G}^{n},\mathcal{O}]]]
\end{equation}
as the n-fold commutation for operators $G$ and $\mathcal{O}$, then in terms of the non-Hermitian Hamiltonian $H^{D}$, $h^{D}$ can be expressed as
\begin{equation}
	h^{D}=-i\theta\left(1+V_{B}\right)+\sum_{n=0}^{\infty}\frac{\left(-1\right)^{n}}{n!}C_{G}^{\left(n\right)}\left(H^{D}\right).
\end{equation}

\indent

The remaining step now, is to find the density matrix of (\ref{hD}), which we will look at in the next section. Consequently, we also obtain the decoherence for our qubit system.

\section{Decoherence of an Anti-$\mathcal{PT}$-symmetric Qubit} 

We shall consider the dynamics where the qubit is initially uncorrelated with a bath in thermal equilibrium i.e. the Gibbs state $\Omega_{B}=e^{-\beta H_{B}}/Z$, where $Z=Tr_{B}e^{-\beta H_{B}}$ is the partition function \cite{Callen1985}, so the initial density matrix of the total system becomes
\begin{equation}
	\rho^{Dh}\left(0\right)=\rho^{Dh}_{S}\left(0\right)\otimes\Omega_{B},
\end{equation}
taking a general form for the qubit
\begin{align}
	\rho^{Dh}_{S}\left(0\right)&=\begin{pmatrix}
							\rho^{Dh}_{11} & \rho^{Dh}_{12} \\
							\rho^{Dh}_{21} & \rho^{Dh}_{21}
							\end{pmatrix} \\
	&=\begin{pmatrix}
	\frac{1}{2}\left(1+\left\langle\sigma_{z}\right\rangle\right) & \left\langle\sigma_{-}\right\rangle \\
	\left\langle\sigma_{+}\right\rangle & \frac{1}{2}\left(1-\left\langle\sigma_{z}\right\rangle\right)
	\end{pmatrix}\,, \label{reducedIC}
\end{align}
with $\sigma_\pm = \sigma_x \pm i\sigma_y$. 
The reduced system's density matrix at time $t$ is given by
\begin{align}
	\rho^{Dh}_{S}\left(t\right)&=\begin{pmatrix}
		\frac{1}{2}\left(1+\left\langle\sigma_{z}\left(t\right)\right\rangle\right) & \left\langle\sigma_{-}\left(t\right)\right\rangle \\
		\left\langle\sigma_{+}\left(t\right)\right\rangle & \frac{1}{2}\left(1-\left\langle\sigma_{z}\left(t\right)\right\rangle\right)
		\end{pmatrix}\label{rhost}
\end{align}
and the expectation value of a general time evolved operator $\mathcal{O}$ with Hermitian Hamiltonian $h^{D}$ can be expressed as $\left\langle\mathcal{O}\left(t\right)\right\rangle=Tr\left[\mathcal{O}\left(t\right)\rho\left(0\right)\right]$ where $\mathcal{O}\left(t\right)=e^{i\int h^{D}dt}\mathcal{O}\left(0\right)e^{-i\int h^{D}dt}$, hence the time-dependent qubit operators are given by
\begin{align}
	\sigma_{z}\left(t\right)&=e^{i\int h^{D}\left(t\right) dt}\sigma_{z}e^{-i\int h^{D}\left(t\right) dt}, \label{sigmazt}\\
	\sigma_{\pm}\left(t\right)&=e^{i\int h^{D}\left(t\right) dt}\sigma_{\pm}e^{-i\int h^{D}\left(t\right) dt} \label{sigmapmt}.
\end{align}

\begin{figure}[h]
	\centering
	\includegraphics[width=0.5\linewidth]{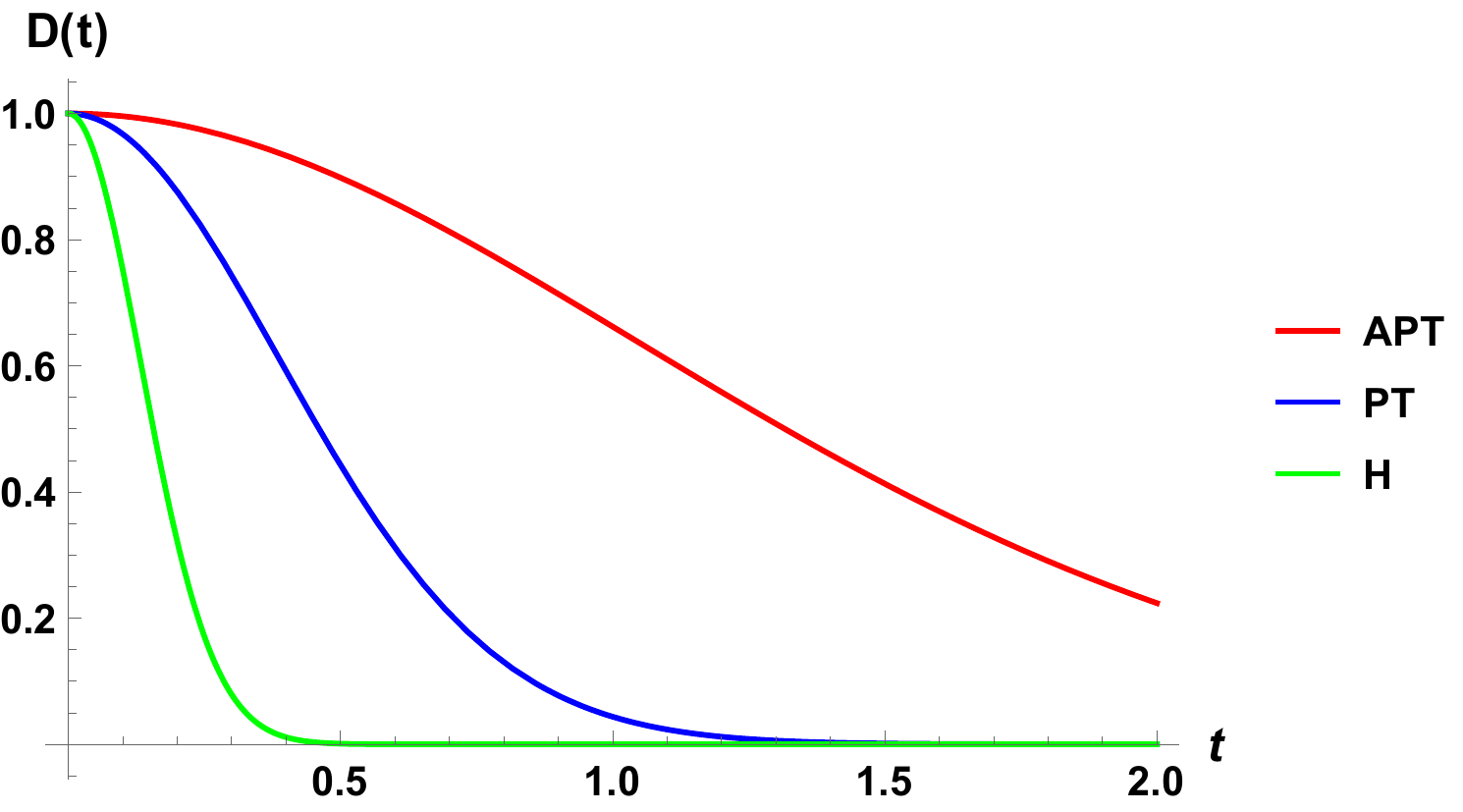}\\
	\caption{Comparison of the evolution of decoherence function with anti-$\mathcal{PT}$-symmetric (APT), $\mathcal{PT}$-symmetric (PT) and Hermitian (H) systems, with parameters $J_{0}=w_{c}=1$, $\beta=0.5$, $\delta=0.56$, $\mu=-0.5$, $\xi=0.81$, $\theta=0.86$ and $\alpha=1$.\label{F1}}
\end{figure}

To proceed with computing the decoherence function, we want to find the expressions for the time-dependent qubit operators (\ref{sigmazt}-\ref{sigmapmt}). To begin, consider the expression for time-dependent bath operator 
\begin{equation}
	a_{k}\left(t\right)=e^{i\int h^{D}\left(t\right) dt}a_{k}e^{-i\int h^{D}\left(t\right) dt} \,, 
\end{equation}
which can be found by first noting the commutation relations $\left[H_{B},a_{k}\right]=-\omega_{k}a_{k}$, $\left[V_{B},a_{k}\right]=-g_{k}$, $\left[\widetilde{V}_{B},a_{k}\right]=-\omega_{k}g_{k}$. 
Then the equation of motion becomes
\begin{equation}
	\frac{d}{dt}a_{k}\left(t\right)=-i\omega_{k}\left[a_{k}\left(t\right)-\omega_{0}\frac{g_{k}}{\omega_{k}}\sigma_{z}+\theta g_{k}t\right],
\end{equation}
and the time-dependent bath operator expression is given by
\begin{equation}
	a_{k}\left(t\right)=-\theta g_{k}t+e^{-i\omega_{k}t}\left[a_{k}-A_{k}\left(t\right)\sigma_{z}+B_{k}\left(t\right)\right],
\end{equation}
where
\begin{equation}
	A_{k}\left(t\right)=\omega_{0}\frac{g_{k}}{\omega_{k}}\left(1-e^{i\omega_{k}t}\right) \quad \text{and} \quad B_{k}\left(t\right)=i\theta\frac{g_{k}}{\omega_{k}}\left(1-e^{i\omega_{k}t}\right).
\end{equation}
Utilizing this expression, we can find the expressions for the time-dependent qubit operators similarly, by using their equations of motion respectively, which are solved by
\begin{align}
	\sigma_{z}\left(t\right) &=\sigma_{z}, \\
	\sigma_{\pm}\left(t\right) & =e^{\mp 2i\omega_{0}t}e^{\int_{0}^{t}\mp 2i\omega_{0}\sum_{k}\left[g_{k}a_{k}^{\dagger}\left(\tau\right)+g_{k}^{\ast}a_{k}\left(\tau\right)\right]d\tau}_{+}\sigma_{\pm},\\
	&=\sigma_{\pm}e^{\mp 2i\omega_{0}t}e^{\pm i\theta\omega_{0}\sum_{k}\left(g^{2}_{k}+\lvert g_{k}\rvert^{2}\right)t^{2}}e^{\mp 4i\theta\omega_{0}\sum_{k}\frac{\lvert g_{k}\rvert^{2}}{\omega_{k}^{2}}\left(1-\cos\left(\omega_{k}t\right)\right)}e^{\pm\sum_{k}\left[2A_{k}\left(t\right)a_{k}^{\dagger}-2A^{\ast}_{k}\left(t\right)a_{k}\right]},\notag
\end{align}
where $e_{+}^{[\cdots]}$ is the time-ordered exponent
\begin{equation}
	e^{2i\omega_{0}\int_{0}^{t}\left[1+\sum G_{k}\left(\tau\right)\right]d\tau}_{+}=e^{2i\omega_{0}\int_{0}^{t}\left[1+\sum G_{k}\left(\tau\right)\right]d\tau}e^{-\omega_{0}\int_{0}^{t}d\tau_{1}\int_{0}^{\tau_{1}}d\tau_{2}\left[G_{k}\left(\tau_{1}\right),G_{k}\left(\tau_{2}\right)\right]},
\end{equation}
with $G_{k}\left(\tau\right)$ a time-dependent operator and $\left[\left[G_{k}\left(\tau_{1}\right),G_{k}\left(\tau_{2}\right)\right],G_{k}\left(\tau_{3}\right)\right]=0$. Now, we can express the reduced system's density matrix at time $t$ (\ref{rhost}), by computing the expectation values of $\sigma_{\pm}\left(t\right)$ and $\sigma_{z}\left(t\right)$
\begin{align}
	\left\langle \sigma_{z}\left(t\right)\right\rangle &=\left\langle \sigma_{z}\right\rangle, \label{sigmazte}\\
	\left\langle \sigma_{\pm}\left(t\right)\right\rangle &=\left\langle \sigma_{\pm}\right\rangle e^{\mp 2i\omega_{0}t}e^{\pm i\omega_{0}\left[\Omega_{2}\left(t\right)-\Omega_{1}\left(t\right)\right]}\left\langle e^{\pm \sum_{k}\left(2A_{k}\left(t\right)a_{k}^{\dagger}-2A_{k}^{\ast}\left(t\right)a_{k}\right)}\right\rangle \label{sigmapmte}\\
	&=\left\langle \sigma_{\pm}\right\rangle e^{\mp 2i\omega_{0}t} e^{\pm i\omega_{0}\left[\Omega_{2}\left(t\right)-\Omega_{1}\left(t\right)\right]}e^{-\omega^{2}_{0}\gamma\left(t\right)},\notag
\end{align}

where
\begin{align}
	\Omega_{1}\left(t\right)&=4\theta\int_{0}^{\infty}dw J\left(w\right)\frac{\left(1-\cos\left(wt\right)\right)}{w^{2}},\\
	\Omega_{2}\left(t\right)&=\theta t^{2}\int_{0}^{\infty}dw \left(J\left(w\right)+\widetilde{J}\left(w\right)\right),
\end{align}
in the continuum limit of bath modes.  Here 
\begin{equation}
	\widetilde{J}\left(w\right)=\sum_{k} \lvert g_{k}\rvert^{2}\left[\cos\left(2\theta_{k}\right)+i\sin\left(2\theta_{k}\right)\right]\delta\left(w-w_{k}\right),
\end{equation}
is the spectral density characterizing the environment with in general, a complex coupling constant $g_{k}=\lvert g_{k}\rvert e^{i\theta_{k}}$. Let us take as an example, $J\left(w\right)=J_{0}w^{1+\mu}e^{-\frac{w}{w_{c}}}$ and $\widetilde{J}\left(w\right)=J\left(w\right)$ with $g_{k}\in\mathbb{R}$. The resulting density matrix of the reduced system at time $t$ is given by

\begin{equation}
	\rho^{Dh}_{S}\left(t\right)=\begin{pmatrix}
						\rho^{Dh}_{11} & 0 \\
						0 & \rho^{Dh}_{22}
						\end{pmatrix}+\begin{pmatrix}
						0 & \rho^{Dh}_{12}e^{ 2i\omega_{0}t}e^{-i\omega_{0}\left[\Omega_{2}\left(t\right)-\Omega_{1}\left(t\right)\right]} \\
						\rho^{Dh}_{21}e^{- 2i\omega_{0}t}e^{i\omega_{0}\left[\Omega_{2}\left(t\right)-\Omega_{1}\left(t\right)\right]} & 0
						\end{pmatrix}e^{-\omega_{0}^{2}\gamma\left(t\right)}\label{AntirhohS}
\end{equation}
and the {\it decoherence function} reads
\begin{equation}
	D\left(t\right)=e^{-\omega_{0}^{2}\gamma\left(t\right)},
\end{equation}
which quantifies the loss of quantum information to the environment. As this function evolves with time, we are able to see that the anti-$\mathcal{PT}$-symmetric (\ref{APTQ}) qubit decays more gradually compared to a Hermitian (\ref{HQ}) or $\mathcal{PT}$-symmetric (\ref{PTQ}) qubit, as shown in Figure \ref{F1}.

\indent

There are four real parameters in our qubits, $\alpha$, $\delta$, $\theta$ and $\xi$. For the Hermitian case, $\alpha$, $\delta$ and $\xi$ will contribute to increasing the decoherence. For a $\mathcal{PT}$-symmetric qubit, the parameters $\delta$, $\xi$ bring an increase and $\theta$, a decrease of decoherence. But for our example of an anti-$\mathcal{PT}$-symmetric qubit, we only have one parameter, $\alpha$ participating in increasing; the other two, $\delta$ and $\xi$, help to reduce decoherence.

\section{Entanglement Entropy} 
Let us now study the entanglement entropy, which is a common measure of entanglement and quantum information. In particular, for our system, this will be a measure of entanglement between our qubit and environment. To do this, let us rewrite the standard representation of the reduced system's density matrix at time $t$ \cite{Allen1975,Cohentannoudji1977} 
\begin{equation}
	\rho^{Dh}_{S}\left(t\right)=\frac{1}{2}\left[1+\vec{v}\left(t\right)\vec{\sigma}\right]
\end{equation}
following \cite{Landau1977} in exponential form

\begin{equation}
	\rho^{Dh}_{S}\left(t\right)=\frac{1}{2}\sqrt{1-v\left(t\right)^{2}}e^{\vec{u}\left(t\right)\cdot\vec{\sigma}},
\end{equation}
where $v\left(t\right)=\left\lvert\vec{v}\left(t\right)\right\rvert$ and $\vec{u}\left(t\right)=\frac{\vec{v}\left(t\right)}{2v\left(t\right)}\ln\left[\frac{1+v\left(t\right)}{1-v\left(t\right)}\right]$. The (von Neumann) entropy can consequently be calculated as
\begin{align}
	S\left(t\right)&=-Tr_{S}\left[\rho^{Dh}_{S}\left(t\right)\ln\rho^{Dh}_{S}\left(t\right)\right],\\
	&=\ln 2-\frac{1}{2}\left[1+v\left(t\right)\right]\ln\left[1+v\left(t\right)\right]-\frac{1}{2}\left[1-v\left(t\right)\right]\ln\left[1-v\left(t\right)\right].\notag
\end{align}
Now taking the initial state of the reduced system (\ref{reducedIC}) as a pure state, we have $Tr\left[\left(\rho^{Dh}_{S}\right)^{2}\left(0\right)\right]=1$, then if we choose equal populations of ground and excited state initially, i.e. $\rho^{Dh}_{11}=\rho^{Dh}_{22}$,
\begin{equation}
	v\left(t\right)=e^{-\omega_{0}^{2}\gamma\left(t\right)}.
\end{equation}

\begin{figure}[H]
	\centering
	\includegraphics[width=0.49\linewidth]{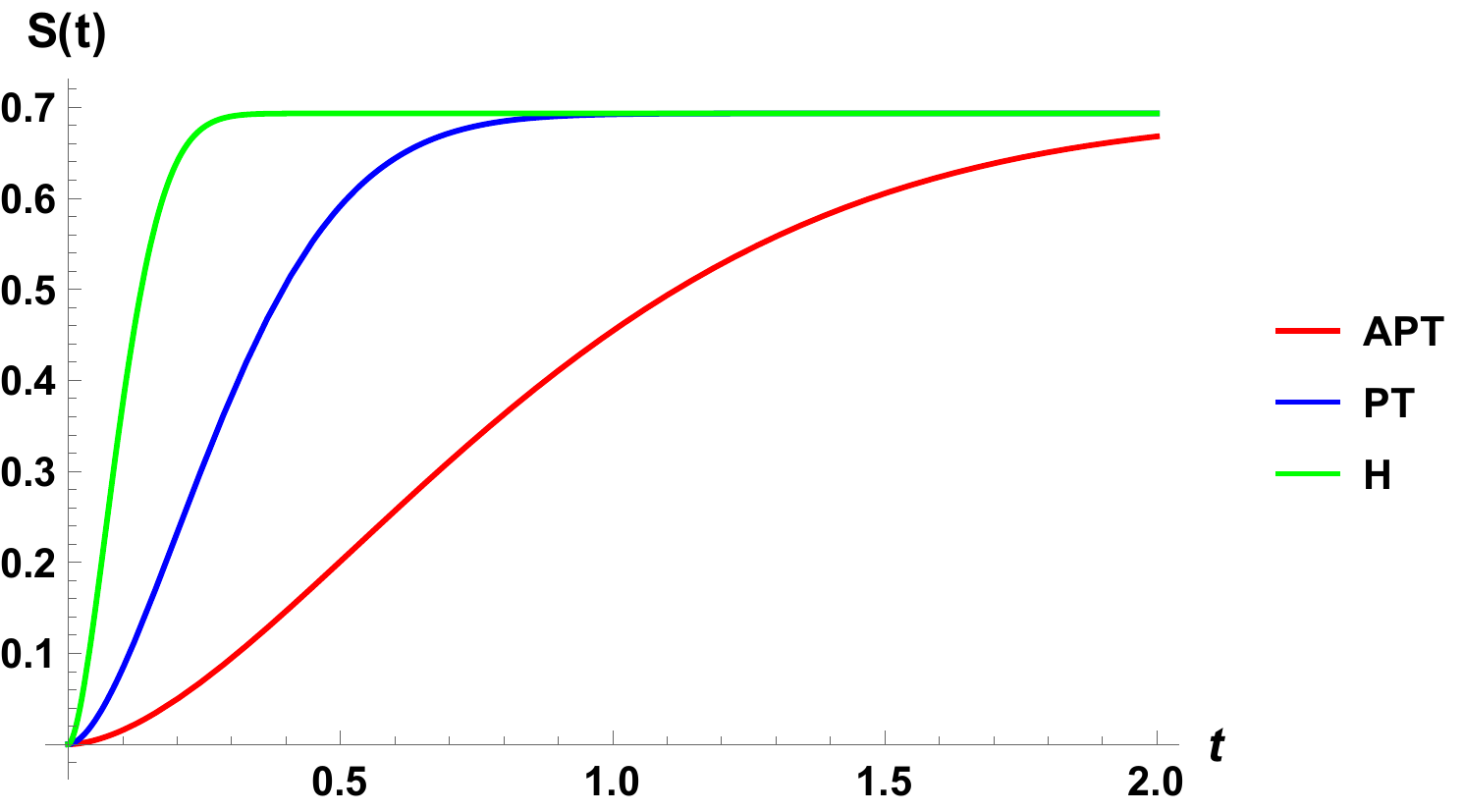}\\
	\caption{Comparison of the time evolution of von Neumann entropy with anti-$\mathcal{PT}$-symmetric (APT), $\mathcal{PT}$-symmetric (PT) and Hermitian (H) systems, with parameters $J_{0}=w_{c}=1$, $\beta=0.5$, $\delta=0.56$, $\mu=-0.5$, $\xi=0.81$, $\theta=0.86$ and $\alpha=1$.\label{F2}}
\end{figure}

The corresponding entropy becomes

\begin{equation}
	S\left(t\right)=\ln 2-\frac{1}{2}\left[1+e^{-\omega_{0}^{2}\gamma\left(t\right)}\right]\ln\left[1+e^{-\omega_{0}^{2}\gamma\left(t\right)}\right]-\frac{1}{2}\left[1-e^{-\omega_{0}^{2}\gamma\left(t\right)}\right]\ln\left[1-e^{-\omega_{0}^{2}\gamma\left(t\right)}\right].
\end{equation}
At $t=0$, $S\left(0\right)=0$ and as $t\rightarrow\infty$, $S\left(\infty\right)=\ln 2$, as shown in Figure \ref{F2}. What is also evident is that the entropy with anti-$\mathcal{PT}$-symmetric qubit increases more gradually compared with Hermitian or $\mathcal{PT}$-symmetric qubit. This result shows the anti-$\mathcal{PT}$-symmetric qubit entangles with the environment at a slower rate compared to Hermitian or $\mathcal{PT}$-symmetric qubits, which is equivalent to the ability to preserve quantum information for a longer time.

An extension of the von Neumann entropy is the quantum R\'{e}nyi entanglement entropy \cite{Renyi1961,MullerLennert2013} through an introduction of a parameter $r$. One interesting connection is the R\'{e}nyi entanglement entropy's close relation with free energy \cite{Baez2011}. In our case, the parameter $r$ gives a {\it nonlinear} dependence of entropy on the reduced density matrix and is given as

\begin{align}
	S_{r}\left(r,t\right)&=\frac{\ln Tr\left\{\left[\rho_{S}^{Dh}\left(t\right)\right]^{r}\right\}}{1-r} \\
	&=\frac{r\ln 2}{1-r}+\frac{1}{1-r}\ln\left\{\left[1+e^{-\omega_{0}^{2}\gamma\left(t\right)}\right]^{r}+\left[1-e^{-\omega_{0}^{2}\gamma\left(t\right)}\right]^{r}\right\}. \notag
\end{align}

\begin{figure}[H]
	\centering
	\includegraphics[width=0.49\linewidth]{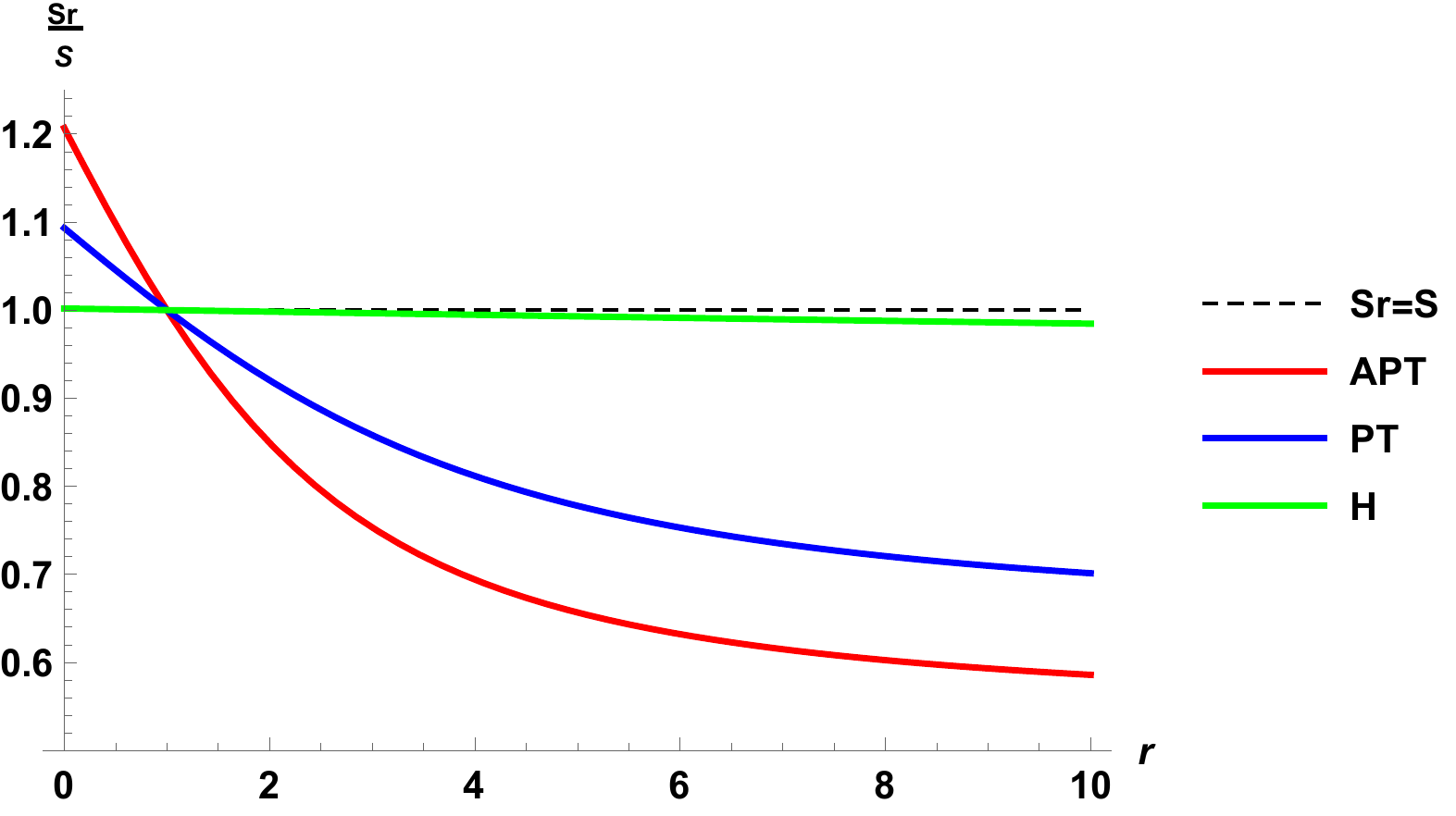}
	\includegraphics[width=0.49\linewidth]{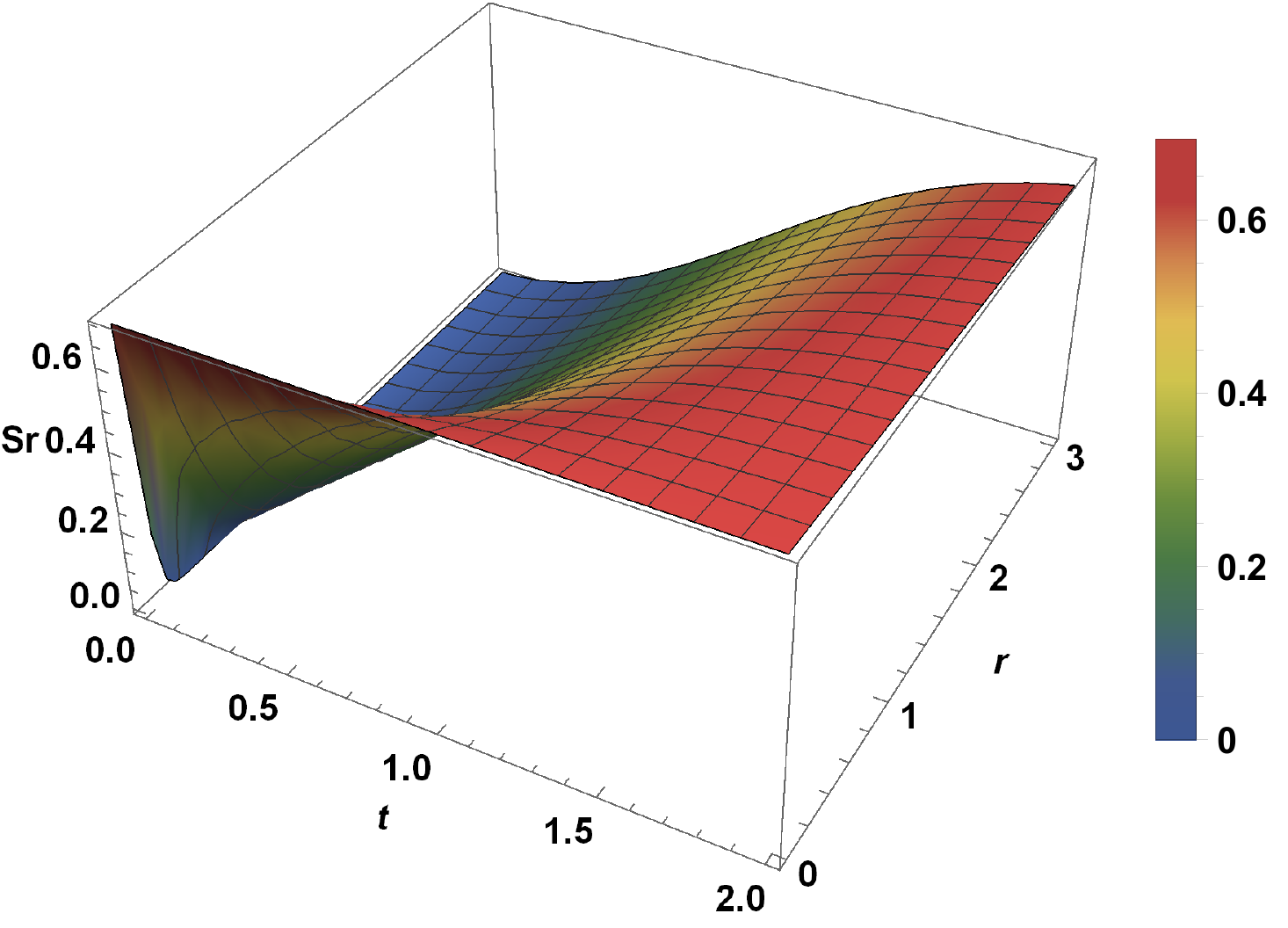}\\
	\caption{Ratio of R\'{e}nyi entanglement entropy and von Neumann entropy at time $1.25$ for anti-$\mathcal{PT}$-symmetric (APT), $\mathcal{PT}$-symmetric (PT) and Hermitian (H) systems (left) and evolution of R\'{e}nyi entanglement entropy for the anti-$\mathcal{PT}$-symmetric case (right), under the parameter choices $J_{0}=w_{c}=1$, $\beta=0.5$, $\delta=0.2$, $\mu=-0.5$, $\xi=0.12$, $\theta=0.05$ and $\alpha=0.3$.\label{F3}}
\end{figure}

Taking the ratio of R\'{e}nyi entanglement entropy with von Neumann entropy at a fixed time for the three types of qubits with respect to  varying the parameter $r$ on the left of Figure \ref{F3}, we can clearly see that in the limit of $r\rightarrow 1$, this reduces back to the von Neumann entropy. On the right of Figure \ref{F3}, we depict the time evolution of R\'{e}nyi entanglement entropy for the anti-$\mathcal{PT}$ qubit under varying parameter $r$.

 To summarize, through analysis of R\'{e}nyi entanglement entropy and von Neumann entropy, results show the anti-$\mathcal{PT}$-symmetric qubit is a better choice to use in quantum computing than the Hermitian or $\mathcal{PT}$-symmetric qubit. We note that this calculation complements the analysis of decoherence function. 

\section{Fisher Information}

Quantum Fisher information is an important quantity in quantum metrology \cite{Toth2013}. It quantifies the precision that can be achieved in estimating a parameter for a given quantum state. Thus, it can 
be regarded as a measure of reliability of a quantum system. A higher value of Fisher information equates to higher precision of estimating a parameter. One can compute from the relative entropy (i.e. the Kullback-Leibler divergence \cite{Kullback1978, Bao2020})

\begin{align}
	D_{KL}\left(K,t\right)&=Tr\left[\rho^{Dh}_{S}\left(\widetilde{K},t\right)\ln \rho^{Dh}_{S}\left(\widetilde{K},t\right)\right]-Tr\left[\rho^{Dh}_{S}\left({K},t\right)\ln \rho^{Dh}_{S}\left(K,t\right)\right],\\
	&=\frac{1}{2}\left\{\ln \frac{1-v^{2}\left(\widetilde{K},t\right)}{1-v^{2}\left(K,t\right)}+v\left(\widetilde{K},t\right)\ln\frac{\left[1+v\left(\widetilde{K},t\right)\right]\left[1-v\left(K,t\right)\right]}{\left[1-v\left(\widetilde{K},t\right)\right]\left[1+v\left(K,t\right)\right]}\right\},\notag
\end{align}

the Fisher entropy with respect to (the inverse) temperature parameter, $K=\beta$ as

\begin{align}
	S_{f}\left(\beta,t\right)&=\frac{\partial^{2}}{\partial \widetilde{\beta}^{2}} D_{KL}\left(\widetilde{\beta},t\right)\bigg\vert_{\widetilde{\beta}=\beta} \\
	&=\frac{\omega_{0}^{4}}{2}\left\{\coth\left[\omega_{0}^{2}\gamma\left(\beta,t\right)\right]-1\right\}\left[\frac{\partial}{\partial\beta}\gamma\left(\beta,t\right)\right]^{2}, \notag
\end{align}

\begin{figure}[H]
	\centering
	\includegraphics[width=0.49\linewidth]{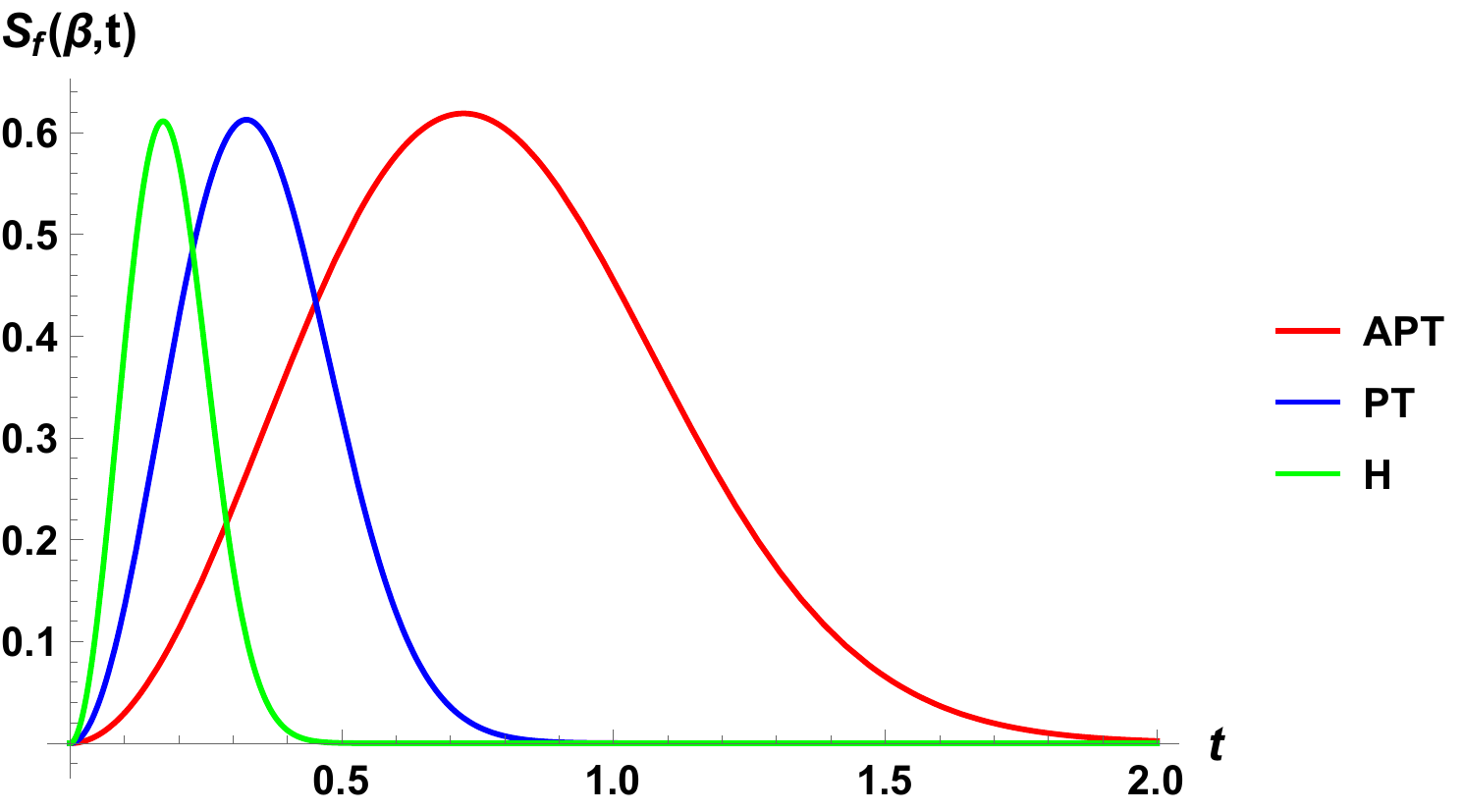}
	\includegraphics[width=0.49\linewidth]{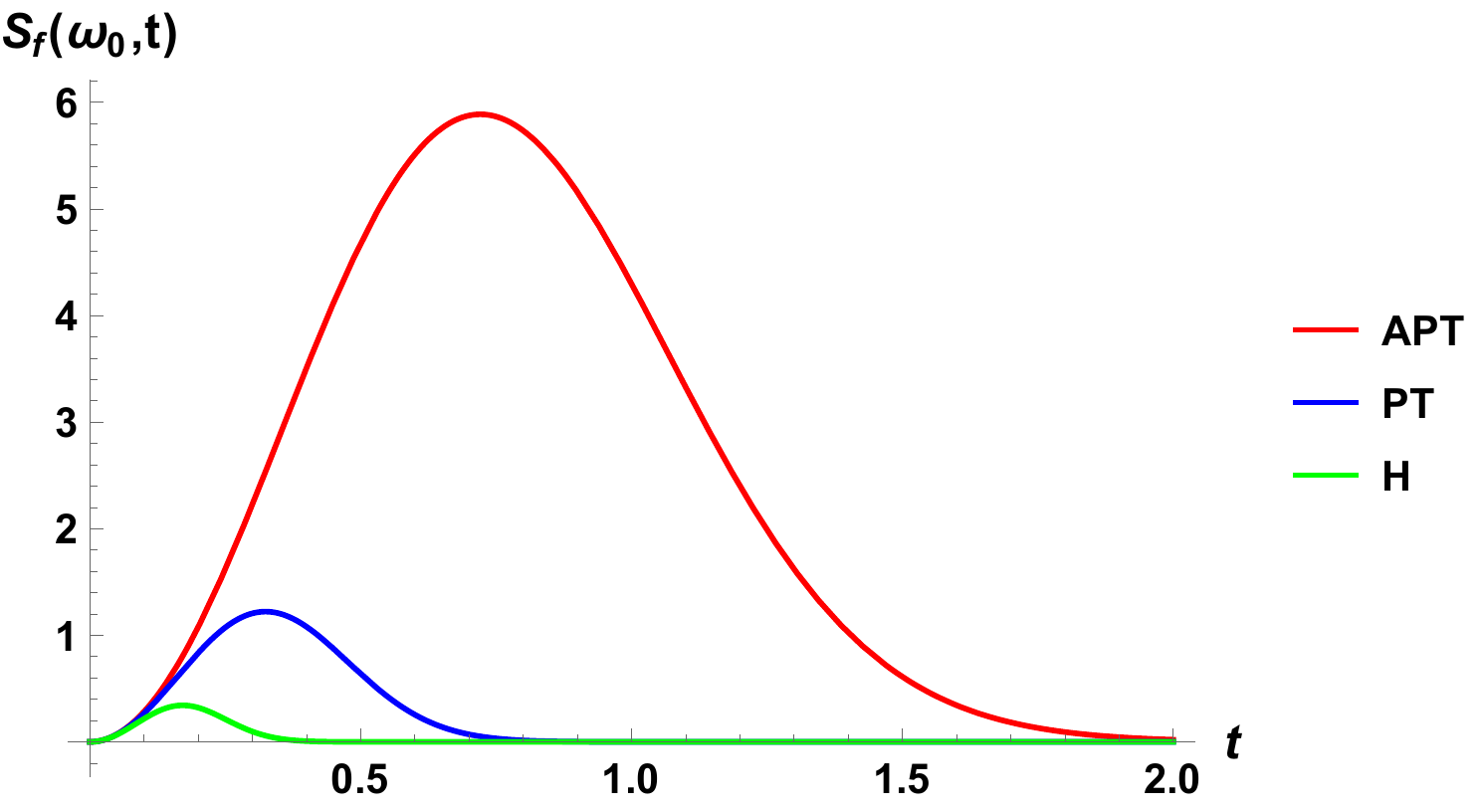}\\
	\caption{Comparison of the Fisher entropy with respect to inverse temperature $\beta$ (left) and with respect to the combined qubit parameter $\omega_{0}$ (right) for anti-$\mathcal{PT}$-symmetric (APT), $\mathcal{PT}$-symmetric (PT) and Hermitian (H) Hamiltonians, with parameters $J_{0}=w_{c}=1$, $\mu=-0.5$, $\beta=\delta=0.5$, $\xi=0.8$, $\theta=0.6$ and $\alpha=1$.\label{F4}}
\end{figure}

From the left of Figure \ref{F4}, we see that although the maximum Fisher information for the three types of qubit are roughly equal, the anti-$\mathcal{PT}$-symmetric case is still slightly higher, as shown in Table \ref{T1}. However, the {\it variance} and area of Fisher information are visibly much larger for the anti-$\mathcal{PT}$-symmetric qubit. This may be interpreted that we need a larger interval of time in estimating $\beta$ accurately.

Similarly, the Fisher entropy depending on what we shall call the  `combined qubit parameter' $\omega_{0}$ can be computed to be
\begin{align}
S_{f}\left(\omega_{0},t\right)&=\frac{\partial^{2}}{\partial \widetilde{\omega_{0}}^{2}} D_{KL}\left(\widetilde{\omega_{0}},t\right)\bigg\vert_{\widetilde{\omega_{0}}=\omega_{0}} \\
&=2\omega_{0}^{2}\left\{\coth\left[\omega_{0}^{2}\gamma\left(\omega_{0},t\right)\right]-1\right\}\gamma^{2}\left(\omega_{0},t\right). \notag
\end{align}

The right of Figure \ref{F4} clearly shows that both the maximum and area of Fisher information with respect to $\omega_{0}$ for the anti-$\mathcal{PT}$-symmetric qubit are much greater compared with the Hermitian and $\mathcal{PT}$-symmetric cases. In particular, this represents that a higher accuracy can be obtained in measuring the parameter $\omega_{0}$. \\ 

In Table \ref{T1}, we compute numerically the maxima, when they occur and total areas of Fisher information.

\begin{table}[H]
	\centering
	\begin{tabular}{|c|c|c||c|c|c||c|}
		\hline
		\multirow{2}{*}{\textbf{Fisher information data}} & \multicolumn{3}{|c|}{\textbf{with parameter $\beta$}} & \multicolumn{3}{|c|}{\textbf{with parameter $\omega_{0}$}} \\ \cline{2-7} 
		 & $S_{f}^{max}$ & $t^{max}$ & $S_{f}^{area}$ & $S_{f}^{max}$ & $t^{max}$ & $S_{f}^{area}$ \\ \hline
		\textbf{Hermitian qubit} &  0.6113  &  0.1714  &  0.1146  &  0.3426  &  0.1713  &  0.0642  \\ \hline
		\textbf{$\mathcal{PT}$-qubit} &  0.6128  &  0.3248  &  0.2192  &  1.2219  &  0.3244  &  0.4366  \\ \hline
		\textbf{Anti-$\mathcal{PT}$-qubit} &  0.6189  &  0.7242  &  0.5058  &  5.8874  &  0.7210  &  4.8039  \\ \hline
	\end{tabular}
\caption{Comparison of the Fisher information data with respect to inverse temperature $\beta$ and the combined qubit parameter $\omega_{0}$ for anti-$\mathcal{PT}$-symmetric, $\mathcal{PT}$-symmetric and Hermitian Hamiltonians, with parameters $J_{0}=w_{c}=1$, $\mu=-0.5$, $\beta=\delta=0.5$, $\xi=0.8$, $\theta=0.6$ and $\alpha=1$.\label{T1}}
\end{table}

From the Table, we can clearly see that the Fisher information maximum and total area for the anti-$\mathcal{PT}$-symmetric qubit are much larger compared with the other two qubits. In addition, the maxima times do not seem to be much affected by the choice of parameters.

\section{Spin Vector Representation}

\begin{figure}[H]
	\centering
	\includegraphics[width=0.45\linewidth]{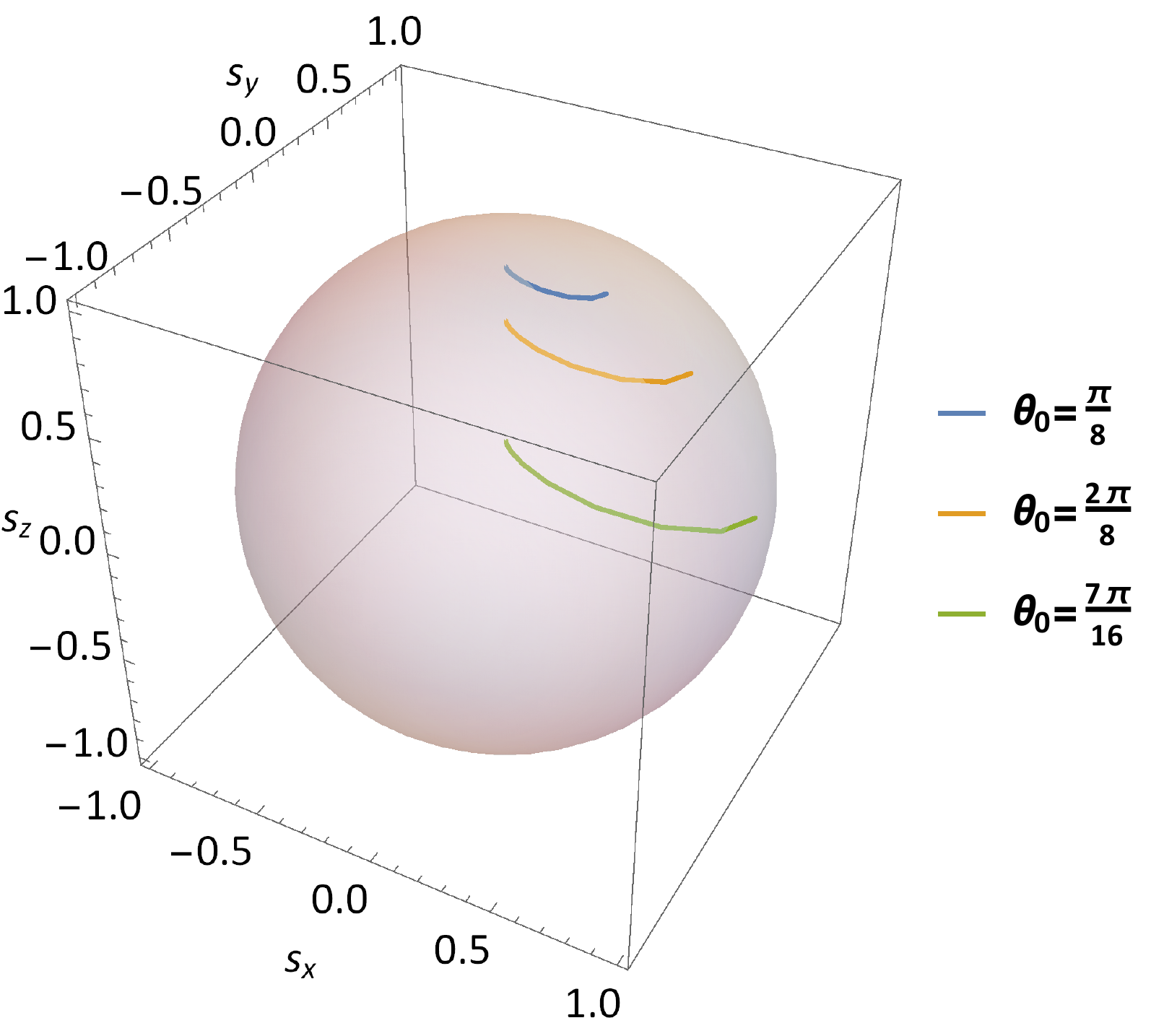}
	\includegraphics[width=0.45\linewidth]{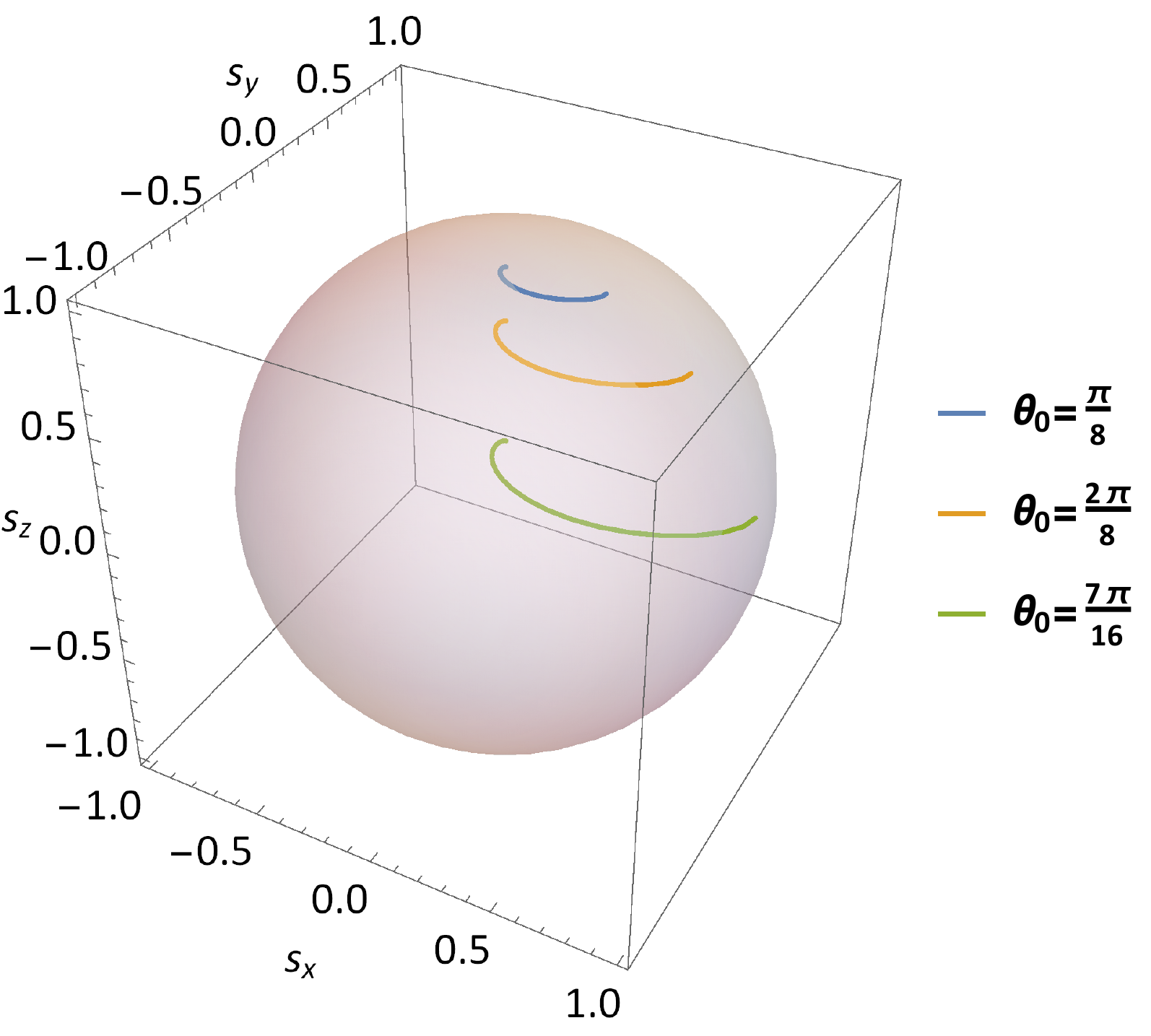}\\
	\includegraphics[width=0.45\linewidth]{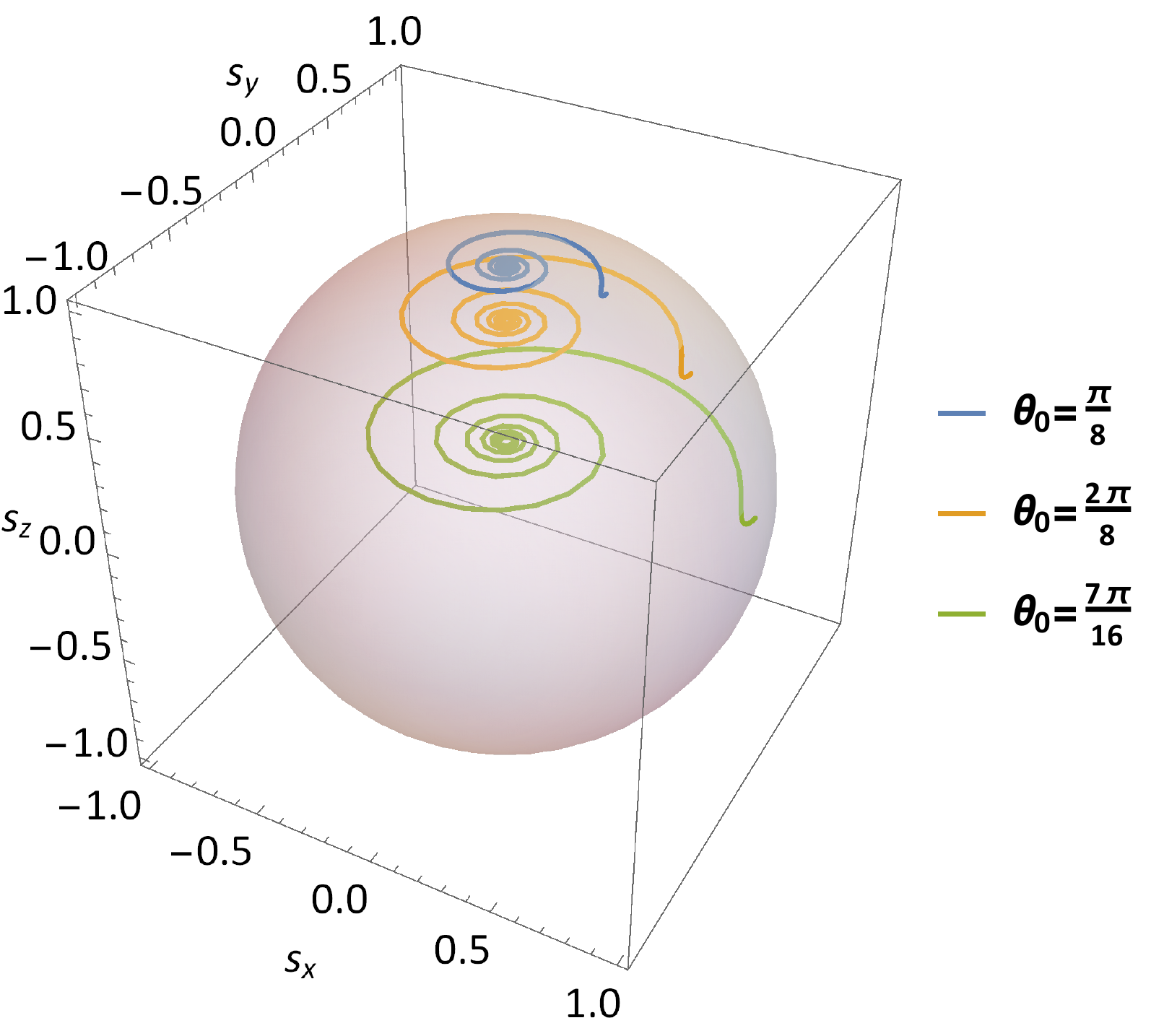}\\
	\caption{Spin vector representation for the evolution of the Hermitian (top left), $\mathcal{PT}$-symmetric (top right) and anti-$\mathcal{PT}$-symmetric (bottom) qubits at different initial $\theta_{0}$, with parameters $J_{0}=w_{c}=1$, $\mu=-0.5$, $\beta=0.5$, $\delta=0.217$, $\xi=0.45$, $\theta=0.488$ and $\alpha=0.5$.\label{F5}}
\end{figure}

\begin{figure}[H]
	\centering
	\includegraphics[width=0.3\linewidth]{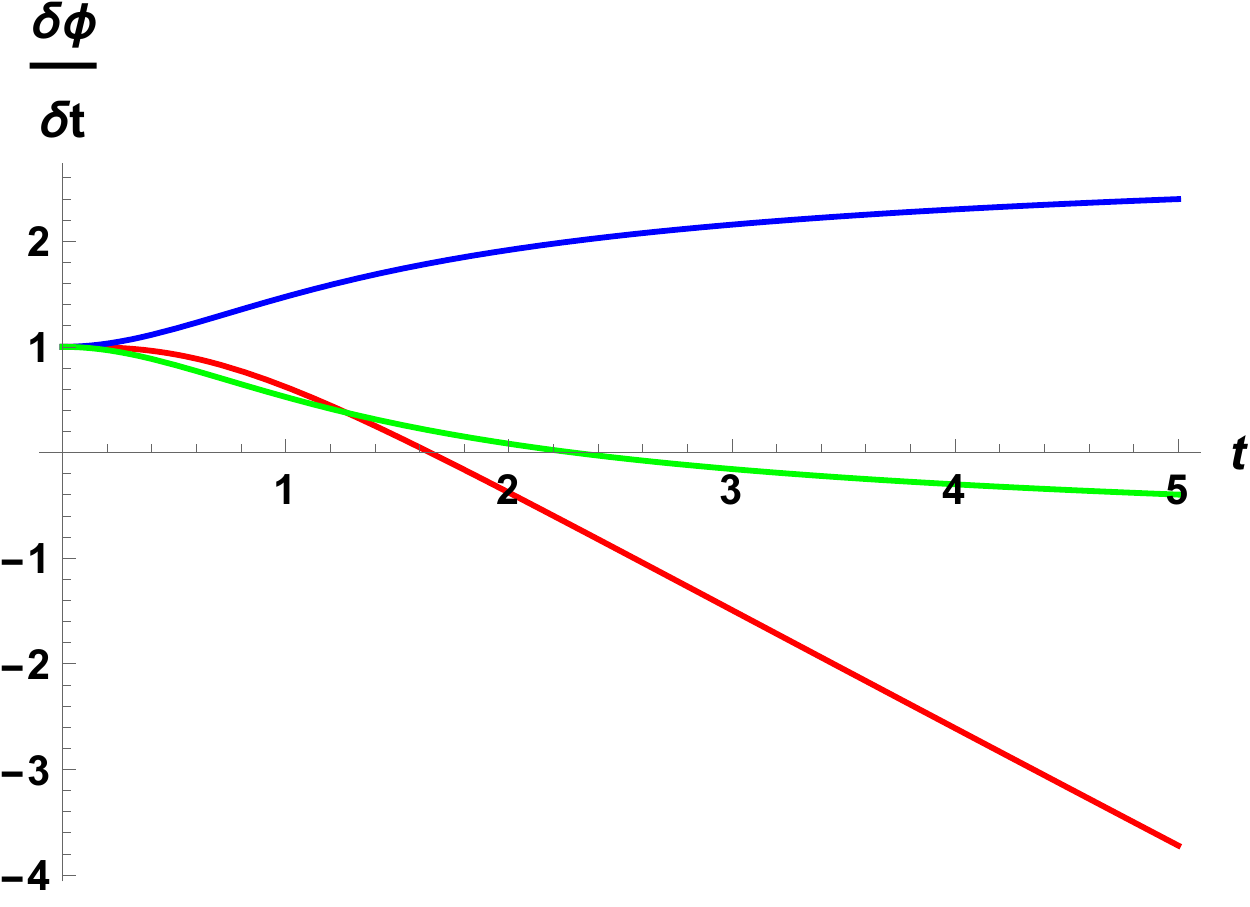} 
	\includegraphics[width=0.3\linewidth]{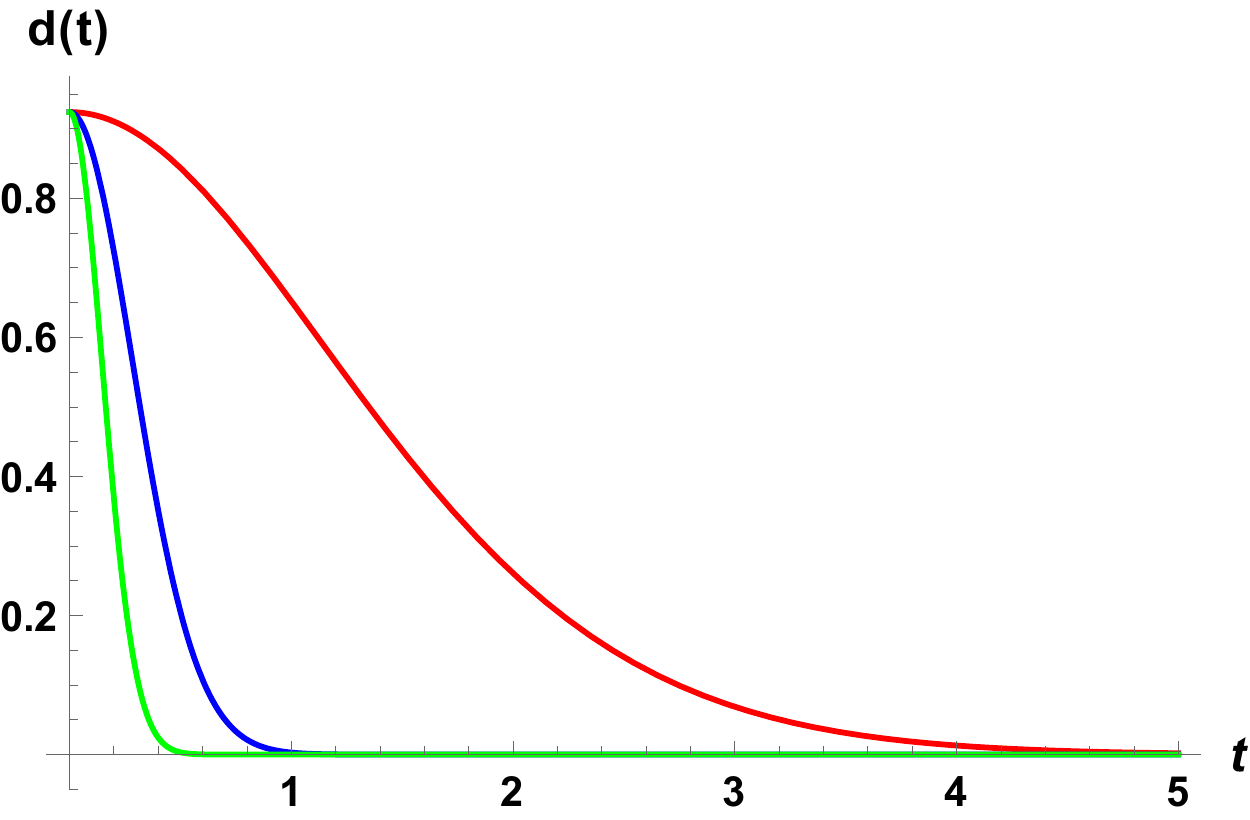} 
	\includegraphics[width=0.38\linewidth]{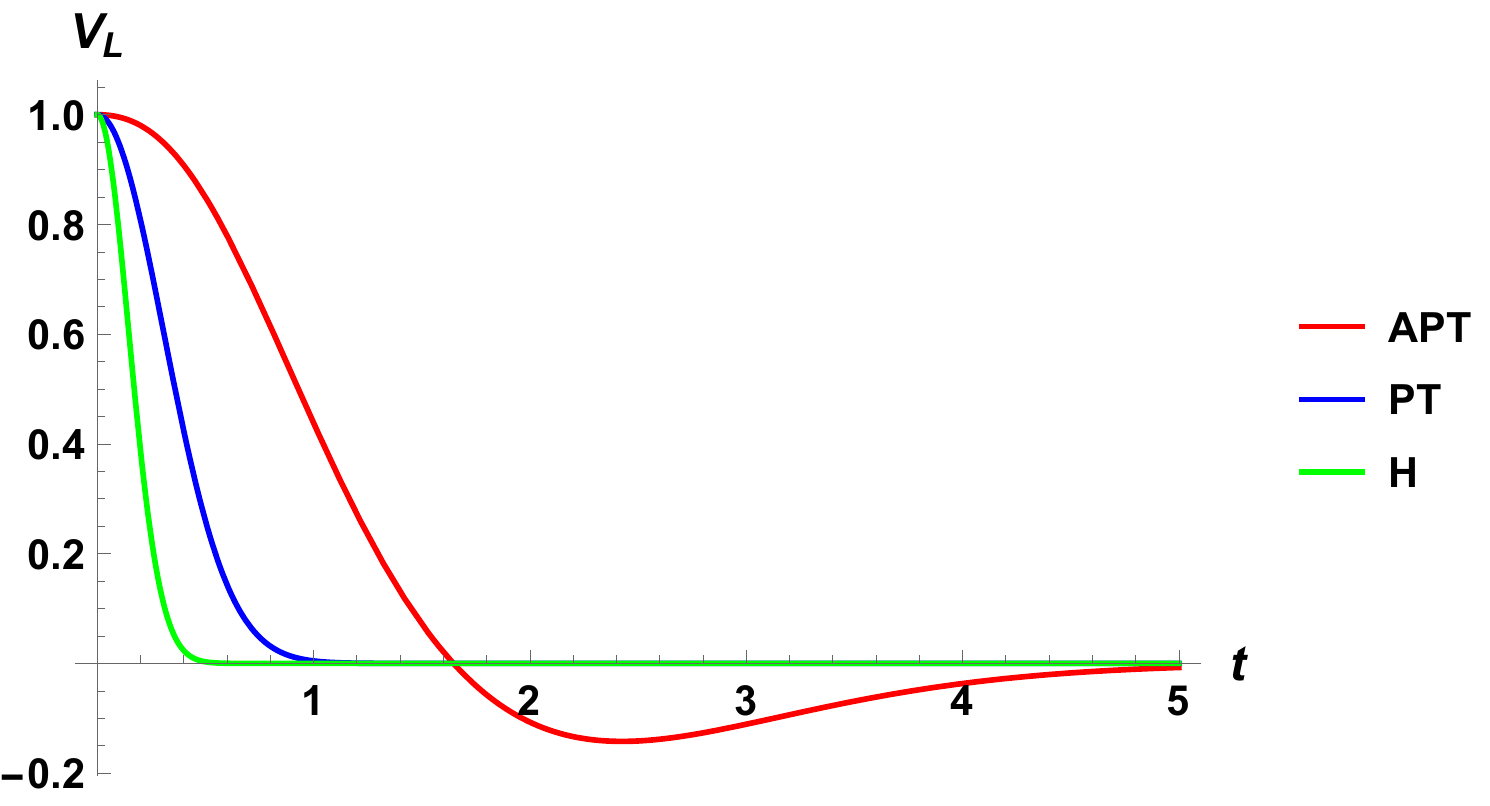}\\
	\caption{Spin angular velocity (left), distance from the $z$-axis (center) and spin linear velocity (right) for the Hermitian (H), $\mathcal{PT}$-symmetric (PT) and anti-$\mathcal{PT}$-symmetric (APT) qubits with parameters $J_{0}=w_{c}=1$, $\mu=-0.5$, $\beta=0.5$, $\delta=0.38$, $\xi=0.8$, $\theta=0.6$ and $\alpha=0.9$ and $\theta_{0}=\frac{3\pi}{8}$.\label{F6}}
\end{figure}

Considering that our investigation is of a two-level system, the dynamics is naturally best revealed with the spin vector representation. Let us start by taking the reduced density matrices of our systems with $g_{K}\in\mathbb{R}$,
\begin{align}
	\rho^{Dh}_{S}\left(t\right)&=\frac{1}{2}\begin{pmatrix}
				1+S_{z} & S_{x}\left(t\right)-iS_{y}\left(t\right) \\
				S_{x}\left(t\right)+iS_{y}\left(t\right) & 1-S_{z}
				\end{pmatrix} \\
				&=\frac{1}{2}\begin{pmatrix}
				1+\cos\theta_{0} & \sin\theta_{0}e^{-i\phi\left(t\right)e^{-\omega_{0}^{2}\gamma\left(t\right)}} \\
				\sin\theta_{0}e^{i\phi\left(t\right)e^{-\omega_{0}^{2}\gamma\left(t\right)}} & 1-\cos\theta_{0}
				\end{pmatrix},\notag
\end{align}
we can obtain the spin vector representation with decoherence as
\begin{align}
	\vec{S}\left(t\right)&=\left(S_{x}\left(t\right),S_{y}\left(t\right),S_{z}\left(t\right)\right) \\
	&=\left(\sin\theta_{0}\cos\left[\phi\left(t\right)\right]D\left(t\right),\sin\theta_{0}\sin\left[\phi\left(t\right)\right]D\left(t\right),\cos\theta_{0}\right),\notag
\end{align}
where the phase is given by 
\begin{equation}
	\phi\left(t\right)=\phi_{0}-2\omega_{0}t+\omega_{0}\widetilde{\Omega}\left(t\right), \label{phase}
\end{equation}
$\omega_{0}$ is the combined qubit parameter
\begin{eqnarray}
	\omega_{0}&=& \left\{\begin{array}{l}
	\sqrt{\alpha^{2}+\delta^{2}+\xi^{2}} \qquad \text{(Hermitian)}\\
	\sqrt{\gamma^{2}+\delta^{2}-\theta^{2}} \qquad \text{($\mathcal{PT}$-symmetric)}\\
	\sqrt{\alpha^{2}-\delta^{2}-\xi^{2}} \qquad \text{(anti-$\mathcal{PT}$-symmetric)}
	\end{array} \right. ,
\end{eqnarray}
and $\widetilde{\Omega}\left(t\right)$ is a time-dependent function in the phase
\begin{eqnarray}
	\widetilde{\Omega}\left(t\right)&=& \left\{\begin{array}{l}
	\Omega\left(t\right) \hspace{0.92in} \text{(Hermitian)}\\
	-\Omega\left(t\right) \hspace{0.79in} \text{($\mathcal{PT}$-symmetric)}\\
	\Omega_{2}\left(t\right)-\Omega_{1}\left(t\right) \hspace{0.26in} \text{(anti-$\mathcal{PT}$-symmetric)}
	\end{array} \right. .
\end{eqnarray}
The dynamics can now be graphically presented against the unit sphere in Figure \ref{F5}, so that we can compare the evolutions of the respective spin vector representation of reduced density matrices for the Hermitian, $\mathcal{PT}$-symmetric and anti-$\mathcal{PT}$-symmetric qubit systems. We take $\phi_{0}=0$ without losing generality and draw the dynamics at different initial angles $\theta_{0}$, with respect to the $S_{z}$ axis. 
 The evolution starts furthest away from the center of the Bloch sphere, then spirals inwards. Initially, all three cases start evolving clockwise, but the Hermitian and anti-$\mathcal{PT}$-symmetric cases change to anticlockwise evolution after a short time. This is visibly noticeable for the anti-$\mathcal{PT}$-symmetric case, but not so easy to see for the Hermitian case, because the speed at which the trajectory reaches the center is much greater than the time it takes to change direction. However, we can take the derivative of phase (\ref{phase}) with respect to time and obtain the {\it spin angular velocity}, which we plot for the three types of qubits after normalization, on the left of Figure \ref{F6}. From the figure, we can see that taking the change of evolution direction clockwise as positive, so when the lines go into the negative region, the trajectories are evolving anticlockwise. What is also noticeable is that the angular speed for the anti-$\mathcal{PT}$-symmetric case seems to increase at a faster rate compared with the other two qubit cases.

\indent

Another feature to note from the spin representation is the distance of evolution trajectory from the $z$-axis
\begin{align}
d\left(t\right)&=\sqrt{S^{2}_{x}\left(t\right)+S^{2}_{y}\left(t\right)} \\
&=\sin\theta_{0}e^{-\omega_{0}^{2}\gamma\left(t\right)},\notag
\end{align}
depicted by the center plot of Figure \ref{F6}. The anti-$\mathcal{PT}$-symmetric distance decreases at a substantially slower rate compared to the other two qubit systems. This plays a large role in the {\it spin linear velocity}, which is given by
\begin{equation}
	V_{L}\left(t\right)=d\left(t\right)\frac{\partial}{\partial t}\phi\left(t\right).
\end{equation}
Drawn against time for the three qubits in the right panel of Figure \ref{F6}, we notice a similar property of a more gradual decay, but for the linear velocity in the anti-$\mathcal{PT}$-symmetric case. 

\section{Conclusions} 
We have studied the decoherence and entanglement (via von Neumann, R\'{e}nyi and Fisher entropy) properties of an anti-$\mathcal{PT}$ qubit comprising a two-level spin system. To this end, we utilized the time-dependent Dyson map to transform our problem of computing the reduced density matrix from a non-Hermitian to a more feasible Hermitian one. We found {\it superior} decoherence properties as compared to the $\mathcal{PT}$-symmetric and Hermitian qubits. We also found a {\it slower} growth for the entanglement entropy and much {\it higher} Fisher information for the anti-$\mathcal{PT}$ qubit. These findings suggest the utility of anti-$\mathcal{PT}$ qubits for quantum information processing and storage. It would be desirable to have a possible experimental realization (e.g. in optical waveguides \cite{Zhang2019} and microcavity systems \cite{Zhang2020}) of the anti-PT-symmetric qubit to observe the predicted superior properties.

\section*{Acknowledgments} We thank Bart\l{}omiej Gardas and Aur\'elia Chenu for insightful discussions. This work was supported by the U.S. Department of Energy.

\end{document}